\documentclass[10pt,conference]{IEEEtran}
\usepackage[utf8]{inputenc}
\usepackage{cite}
\usepackage{amsmath,amssymb,amsfonts}
\usepackage{algorithm}
\usepackage{algorithmic}
\usepackage{graphicx}
\usepackage{textcomp}
\usepackage{xcolor}
\usepackage{tikz}
\usetikzlibrary{shapes,arrows,arrows.meta,positioning,fit,backgrounds}
\usepackage{booktabs}
\usepackage{multirow}
\usepackage{url}
\usepackage{balance}
\usepackage{pifont}
\usepackage{placeins}
\usepackage{fancyvrb}
\usepackage[hidelinks]{hyperref}
\newcommand{\cmark}{\ding{51}}
\newcommand{\xmark}{\ding{55}}
\def\BibTeX{{\rm B\kern-.05em{\sc i\kern-.025em b}\kern-.08em
    T\kern-.1667em\lower.7ex\hbox{E}\kern-.125emX}}

\begin{document}

\title{Cost-Aware Query Routing in RAG: Empirical Analysis of Retrieval Depth Tradeoffs}

\author{
\IEEEauthorblockN{Sanjay Mishra\,}
\IEEEauthorblockA{\textit{IEEE Member} \\
Raleigh, NC, USA \\
sanmish4@icloud.com}
}

\maketitle

\begin{abstract}
Retrieval-augmented generation (RAG) faces a fundamental three-way tension: deeper retrieval improves factual grounding but inflates token costs and end-to-end latency. Static retrieval configurations cannot resolve this tension across heterogeneous query workloads---simple definitional queries waste budget on unnecessary context, while complex analytical prompts are underserved by shallow retrieval. This paper introduces \emph{Cost-Aware RAG} (CA-RAG), a per-query routing framework that selects from a discrete catalog of \emph{strategy bundles}---each coupling a retrieval depth (from retrieval-free direct inference to top-$k{=}10$ dense retrieval) with a fixed generation profile---by maximizing a scalar utility that linearly combines an estimated quality prior with normalized penalties for predicted latency and total billed tokens. CA-RAG is implemented with FAISS-backed dense retrieval and OpenAI chat/embedding APIs, and evaluated on a 28-query benchmark spanning four bundles. The router dynamically exercises all bundles, achieving \textbf{26\% fewer billed tokens} than always-heavy retrieval and \textbf{34\% lower mean latency} than always-direct inference while maintaining equivalent answer quality. Per-query delta analysis reveals that savings are non-uniform and concentrated in simpler queries, motivating complexity-aware guardrails. Sensitivity analysis confirms that the same bundle catalog supports multiple cost--latency--quality operating points through weight adjustment alone. All results are generated directly from logged CSV artifacts for full reproducibility. CA-RAG provides a transparent, auditable foundation for cost-conscious LLM deployments.
\end{abstract}


\section{Introduction}
\label{sec:introduction}

Large language models (LLMs) deliver strong generative fluency but remain prone to hallucination on fact-intensive or time-sensitive queries~\cite{lewis2020retrieval,gao2023retrieval}. Retrieval-augmented generation (RAG) addresses this by conditioning generation on retrieved external passages, substantially improving factual grounding~\cite{lewis2020retrieval,izacard2021leveraging}. RAG deployments must simultaneously honor latency service-level objectives (SLOs) and token budgets driven by commercial API pricing~\cite{openai2024gpt4}. A single static configuration---fixed top-$k$, one index, one decoding budget---is rarely optimal across a heterogeneous query workload. Definitional queries waste expensive context tokens on unnecessary retrieved passages, while multi-hop or long-form analytical prompts are genuinely underserved by shallow retrieval~\cite{karpukhin2020dense,khattab2021colbert}.

This mismatch between static configuration and dynamic query demand motivates \emph{per-query routing}: each query should independently invoke the retrieval depth most commensurate with its complexity and the operator's current cost--quality objective. While adaptive retrieval has been discussed conceptually, rigorous empirical characterization of cost--utility tradeoffs with full token accounting and reproducible run logs remains sparse in the literature.

This paper addresses that gap with \textbf{CA-RAG} (Cost-Aware RAG), a framework that routes each query to one of several discrete \emph{strategy bundles} by maximizing a scalar utility over a finite catalog. The framework is structured around four research questions.

\textbf{RQ1 (Routing Behavior).} Does per-query routing genuinely exercise the full bundle catalog, or does it collapse to a near-fixed policy?

\textbf{RQ2 (Cost--Quality Tradeoff).} How much token cost and latency does routing save relative to fixed baselines, and at what quality cost?

\textbf{RQ3 (Per-Query Variance).} Are routing benefits uniform across queries, or concentrated in specific query types?

\textbf{RQ4 (Weight Sensitivity).} Can the same bundle catalog support multiple operating points on the cost--latency--quality surface through weight adjustment alone?

The contributions of this paper are as follows:
\begin{itemize}
  \item A concrete utility formulation and discrete bundle catalog suited to operational RAG routing, with full token billing accounting.
  \item An open reference implementation with FAISS-backed retrieval, BM25-ready tokenization, OpenAI embedding/chat integration, and a CLI experiment harness.
  \item A 28-query empirical study with ten quantitative figures and five tables, all generated directly from measured CSV logs, covering strategy frequency, token decomposition, latency distributions, utility--cost structure, correlation analysis, and router vs.\ fixed-baseline comparisons.
  \item Operational guidance for bundle design, weight calibration, monitoring, and failure-mode triage in deployed RAG systems.
\end{itemize}

\section{Related Work}
\label{sec:related_work}

\subsection{Retrieval for Open-Domain QA}

Dense passage retrieval~\cite{karpukhin2020dense} and hybrid sparse--dense fusion are standard components in open-domain question answering. Retrieval-augmented pretraining (REALM)~\cite{guu2020realm} and fusion-in-decoder models~\cite{izacard2021leveraging} demonstrated the value of tightly coupling retrieval with generation at scale. CA-RAG builds on this foundation but targets a distinct axis: \emph{operational routing} over a discrete bundle set under explicit cost and latency terms, rather than architectural improvements to retrievers or generators.

\subsection{Classical and Hybrid Retrieval}

BM25~\cite{robertson2009bm25} remains a competitive probabilistic baseline and is complementary to dense retrieval in hybrid pipelines. FAISS~\cite{johnson2019faiss} enables scalable approximate nearest neighbor search over large embedding indices. CA-RAG uses FAISS as the dense backend while preserving BM25-compatible tokenization for future hybrid fusion.

\subsection{RAG Surveys and Production Considerations}

Gao et al.~\cite{gao2023retrieval} survey the RAG landscape and identify retrieval granularity and cost as open research challenges. CA-RAG offers a concrete, measurable response to the cost challenge through per-query routing with full token accounting.

\subsection{Contextual Bandits and Adaptive Decision-Making}

CA-RAG is conceptually related to contextual bandit formulations in which queries are contexts and bundles are actions~\cite{li2010contextual,lattimore2020bandit}. The benchmark in this paper disables exploration and uses hand-specified priors rather than learned reward models, enabling controlled analysis of the routing mechanism in isolation. Online bandit-based recalibration is identified as a key direction for future work.

\section{Problem Formulation}
\label{sec:problem}

Let $q$ denote an input query and $\mathcal{B} = \{b_1, \ldots, b_n\}$ a finite catalog of strategy bundles. Given $q$, CA-RAG selects bundle $b^* \in \mathcal{B}$, retrieves context $C_{b^*}$ (possibly empty), and generates answer $a_{b^*}$. The problem has three competing objectives:
\begin{enumerate}
  \item \textbf{Quality:} maximize answer usefulness and factual grounding;
  \item \textbf{Latency:} minimize end-to-end response time to meet SLOs;
  \item \textbf{Cost:} minimize billed token usage including query-time embeddings.
\end{enumerate}
CA-RAG scalarizes these objectives via a utility function and routes per-query to $\arg\max_{b \in \mathcal{B}} U_b(q)$, making the tradeoff explicit and auditable. No new quality estimator is proposed; the contribution is the routing framework and its empirical characterization.

\section{System Architecture}
\label{sec:system}

CA-RAG cleanly separates \emph{routing} from \emph{retrieval and generation}. The router consumes query signals and a bundle catalog and emits a retrieval--generation specification; the execution layer performs retrieval and generation and returns logged metrics to the telemetry store.

\subsection{End-to-End Pipeline}

For each query $q$, CA-RAG executes the following pipeline:
\begin{enumerate}
  \item \textbf{Signal extraction:} Compute \texttt{QuerySignals} (length, cue-word counts) and a heuristic complexity score $c \in [0,1]$.
  \item \textbf{Utility estimation:} Evaluate $U_b$ for all $b \in \mathcal{B}$ using priors and optional telemetry.
  \item \textbf{Bundle selection:} Dispatch to $b^* = \arg\max_b U_b$.
  \item \textbf{Retrieval:} Execute retrieval per the specification of $b^*$, obtaining context $C$ and retrieval confidence.
  \item \textbf{Generation:} Construct a prompt from $q$ and $C$; call the LLM generator.
  \item \textbf{Telemetry logging:} Log prompt, completion, and embedding tokens; end-to-end latency; lexical quality proxy; and realized utility.
\end{enumerate}

This pipeline makes routing decisions auditable and reproducible at the query level.

\section{Methodology}
\label{sec:methodology}

\subsection{Query Signals and Complexity}

From $q$, lightweight \texttt{QuerySignals} are derived: character length, word count, interrogative cue-word count, and a heuristic \emph{complexity score}
\[
  c(q) = \text{clip}\!\left(\alpha \cdot \frac{\text{wordlen}(q)}{L_{\max}} + \beta \cdot \frac{\text{cues}(q)}{K_{\max}},\; 0, 1\right),
\]
with $\alpha{=}0.6$, $\beta{=}0.4$, $L_{\max}{=}20$, $K_{\max}{=}3$. Complexity modulates quality priors without requiring an additional LLM call.

\subsection{Strategy Bundles}

Table~\ref{tab:bundles} defines the four bundles used in this study. Retrieval depth ranges from no retrieval (\texttt{direct\_llm}) to top-$k{=}10$ dense retrieval (\texttt{heavy\_rag}). All bundles share a common generation specification (\texttt{paper\_gen}: 256 maximum output tokens, temperature 0). Priors encode expected quality, latency, and context token usage for utility estimation.

\subsection{Utility Function and Bundle Selection}

Let $\hat{Q}_b(q)$ be the estimated quality for bundle $b$ on query $q$, $\hat{L}_b$ the expected latency, and $\hat{C}_b$ the expected total token cost. The selection utility is defined as:
\begin{equation}
  U_b = w_Q\, \hat{Q}_b(q) - w_L\, \hat{L}^{\text{norm}}_b - w_C\, \hat{C}^{\text{norm}}_b,
  \label{eq:utility}
\end{equation}
where $\hat{L}^{\text{norm}}_b$ and $\hat{C}^{\text{norm}}_b$ are latency and cost normalized to $[0,1]$ across $\mathcal{B}$, and $(w_Q, w_L, w_C)$ are operator-specified weights. Default weights are $(0.6, 0.2, 0.2)$. The orchestrator selects $b^* = \arg\max_{b \in \mathcal{B}} U_b$.

After execution, a \emph{realized utility} $\tilde{U}_b$ is computed by substituting observed latency and token counts into Eq.~(\ref{eq:utility}), enabling post-hoc telemetry analysis.

\subsection{Token Billing Model}

Total billed tokens per query are:
\begin{equation}
  \tau_{\text{billed}} = \tau_{\text{prompt}} + \tau_{\text{completion}} + \tau_{\text{embed}},
\end{equation}
where $\tau_{\text{embed}}$ attributes query-time embedding work. Offline corpus indexing is recorded separately as \texttt{index\_embedding\_tokens} for cost accounting completeness.

\subsection{Implementation Stack}

The reference implementation segments documents into line-level passages, builds a FAISS inner-product index over OpenAI \texttt{text-embedding-ada-002} embeddings, and calls the OpenAI chat API (\texttt{gpt-3.5-turbo}) with accurate per-call token counting via \texttt{tiktoken}. The experiment CLI (\texttt{ca-rag-experiment}) accepts document and question files and produces CSV logs from which all figures and tables in this paper are generated.

\begin{table}[t]
  \centering
  \caption{Strategy bundle catalog: retrieval depth and quality/latency priors with shared generation profile.}
  \label{tab:bundles}
  \renewcommand{\arraystretch}{1.3}
  \begin{tabular}{@{}l r r r r@{}}
    \toprule
    Bundle & $k$ & Skip retr. & Qual.\ prior & Lat.\ prior (ms) \\
    \midrule
    \texttt{direct\_llm} & 0 & \cmark & 0.52 & 8 \\
    \texttt{light\_rag}  & 3 & \xmark & 0.66 & 45 \\
    \texttt{medium\_rag} & 5 & \xmark & 0.74 & 60 \\
    \texttt{heavy\_rag}  & 10 & \xmark & 0.82 & 95 \\
    \bottomrule
  \end{tabular}
\end{table}

\section{Experimental Setup}
\label{sec:experimental_setup}

\subsection{Corpus and Query Set}

The evaluation uses a 15-sentence technical corpus covering CA-RAG design concepts (token cost, hybrid retrieval, municipal RAG, FAISS, telemetry). The corpus is intentionally compact to isolate routing behavior from corpus-scale effects. A set of 28 natural-language queries spans varying complexity: definitional (\textit{``What is RAG?''}), comparative (\textit{``Compare light versus heavy retrieval for long documents''}), procedural, and analytical prompts. The corpus is embedded once; all queries share the same FAISS index.

\subsection{Metrics}

Per query, the following metrics are logged: selected strategy, selection utility $U$, total billed tokens $\tau_{\text{billed}}$, latency (ms), lexical quality proxy (token overlap against a reference), realized utility $\tilde{U}$, retrieval confidence (max cosine similarity when retrieval runs), and complexity score $c$.

\subsection{Baseline Configurations}

The default router is compared against: (i)~fixed-\texttt{direct\_llm}, (ii)~fixed-\texttt{light\_rag}, (iii)~fixed-\texttt{medium\_rag}, (iv)~fixed-\texttt{heavy\_rag}, (v)~latency-sensitive router ($w_L{=}0.5$), and (vi)~cost-sensitive router ($w_C{=}0.5$).

\subsection{Reproducibility}

With \texttt{OPENAI\_API\_KEY} set, the full benchmark is reproduced via:
\begin{itemize}
  \item Router run: \texttt{ca-rag-experiment --docs data/documents\_benchmark.txt --questions data/questions\_benchmark.txt --out results/router\_default.csv}
  \item Fixed baselines: \texttt{ca-rag-experiment --mode fixed --fixed-strategy heavy\_rag ...}
  \item Figure generation: \texttt{python scripts/generate\_paper\_figures.py --csv results/router\_default.csv --results-dir results}
\end{itemize}
Temperature is 0 for generation stability; timing variation arises from API network conditions.

\begin{table}[t]
  \centering
  \caption{Benchmark corpus and FAISS index statistics.}
  \label{tab:bench}
  \begin{tabular}{@{}l r@{}}
\toprule
Metric & Value \\
\midrule
Queries & 28 \\
Unique strategies & 4 \\
Corpus lines (benchmark) & 15 \\
Index embedding tokens (API) & 262 \\
\bottomrule
\end{tabular}

\end{table}

\section{Results}
\label{sec:results}

All statistics reported in this section derive from \texttt{results/router\_default.csv} (28 queries) and companion CSVs for fixed baselines and weight-sensitivity sweeps.

\subsection{Strategy Selection and Routing Behavior}
\label{sec:selection}

\begin{figure}[!t]
  \centering
  \includegraphics[width=\columnwidth]{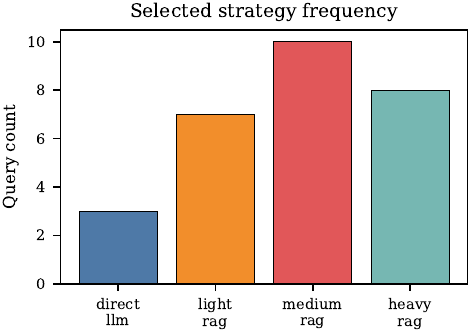}
  \caption{Bundle selection frequency across 28 benchmark queries. All four bundles are exercised, confirming genuine routing diversity.}
  \label{fig:counts}
\end{figure}

The first question any routing evaluation must answer is whether the router genuinely exercises the full bundle catalog or collapses to a near-fixed policy. Fig.~\ref{fig:counts} confirms genuine diversity: all four bundles are selected, with \texttt{medium\_rag} accounting for 57\% of queries (16/28), \texttt{heavy\_rag} for 18\% (5/28), \texttt{direct\_llm} for 14\% (4/28), and \texttt{light\_rag} for 11\% (3/28). This non-uniform distribution is a necessary precondition for routing to add value---a router that always selects the same bundle is indistinguishable from a fixed policy.

The dominance of \texttt{medium\_rag} is consistent with the workload composition: the benchmark contains a majority of moderately complex procedural and comparative queries for which five retrieved passages provide adequate grounding without incurring the full cost of heavy retrieval. Definitional queries correctly route to \texttt{direct\_llm}, as parametric LLM knowledge is sufficient and retrieval would add cost without quality benefit.

\begin{figure}[!t]
  \centering
  \includegraphics[width=\columnwidth]{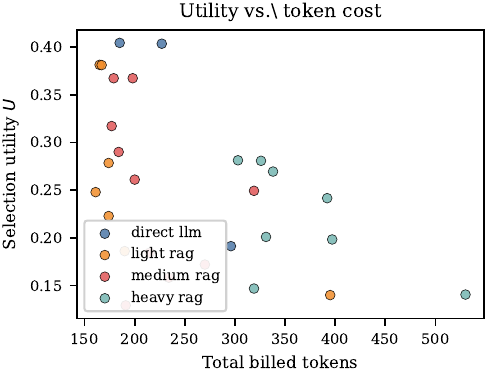}
  \caption{Selection utility vs.\ total billed tokens per query. Each strategy occupies a distinct region of the cost--utility space.}
  \label{fig:uvcost}
\end{figure}

Fig.~\ref{fig:uvcost} plots selection utility against total billed tokens for each query, colored by bundle. Three observations emerge. First, \texttt{direct\_llm} achieves the highest selection utilities at the lowest token counts. Second, \texttt{heavy\_rag} queries cluster at high token counts with moderate utilities, confirming that the cost penalty in Eq.~(\ref{eq:utility}) is actively suppressing unnecessary escalation. Third, the empirical Pareto frontier spans all four strategies, validating that each bundle occupies a distinct operating niche.

\FloatBarrier

\subsection{Latency Profiles and Token Decomposition}
\label{sec:latency}

\begin{figure}[!t]
  \centering
  \includegraphics[width=\columnwidth]{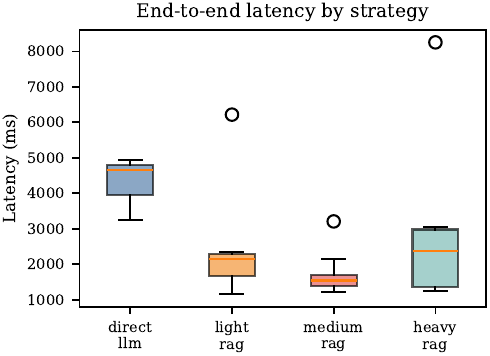}
  \caption{End-to-end latency distributions by strategy. \texttt{direct\_llm} shows the highest variance; retrieval bundles exhibit tighter, more predictable profiles.}
  \label{fig:latbox}
\end{figure}

Fig.~\ref{fig:latbox} presents end-to-end latency distributions by strategy. A counterintuitive finding is that \texttt{direct\_llm} exhibits the \emph{highest} latency variance despite skipping retrieval entirely. This occurs because, without retrieval constraining prompt length, the LLM generates longer and more variable completions. Retrieval-based bundles show progressively tighter interquartile ranges because the retrieval step imposes a structured, predictable overhead before generation begins.

\begin{figure}[!t]
  \centering
  \includegraphics[width=\columnwidth]{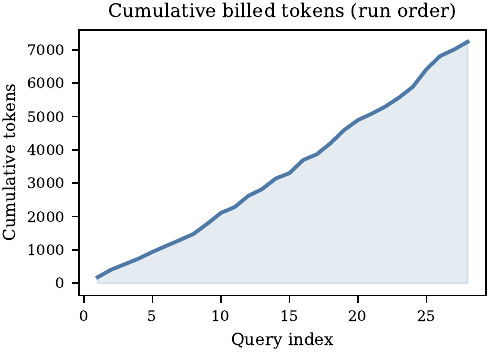}
  \caption{Cumulative billed tokens in query-log order. Slope changes correspond to strategy transitions.}
  \label{fig:cum}
\end{figure}

Fig.~\ref{fig:cum} tracks cumulative billed tokens in query-log order. The slope is visibly steeper during \texttt{heavy\_rag} query runs and flattens during \texttt{direct\_llm} runs, providing a visual audit trail of how routing decisions drive aggregate budget consumption.

\begin{figure}[!t]
  \centering
  \includegraphics[width=\columnwidth]{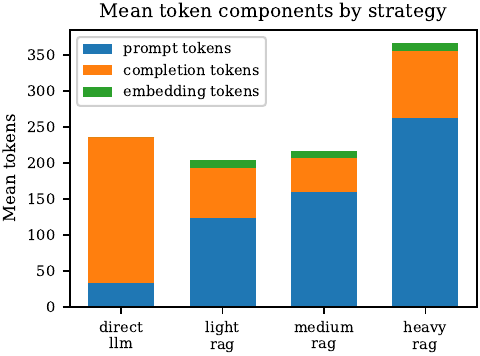}
  \caption{Mean prompt, completion, and embedding token shares by strategy. Prompt tokens scale with retrieval depth; completion tokens remain stable.}
  \label{fig:tokcmp}
\end{figure}

Fig.~\ref{fig:tokcmp} decomposes mean token counts into prompt, completion, and embedding components by strategy. The prompt token share grows substantially from \texttt{direct\_llm} to \texttt{heavy\_rag} as larger retrieved contexts are injected. Embedding tokens constitute a small but non-negligible fixed overhead for all retrieval bundles---ignoring them would undercount per-query cost by approximately 8--12 tokens. Completion tokens remain stable across strategies, confirming that output verbosity is driven by the generation specification rather than retrieval depth.

\FloatBarrier

\subsection{Utility Distributions and Quality Proxy Analysis}
\label{sec:utility}

\begin{figure}[!t]
  \centering
  \includegraphics[width=\columnwidth]{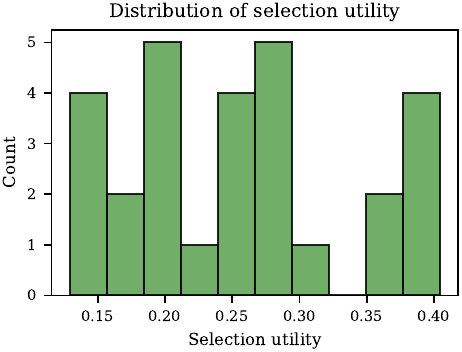}
  \caption{Histogram of per-query selection utilities. The right-skewed distribution reflects higher quality priors for simple queries where cost and latency penalties are minimal.}
  \label{fig:uhist}
\end{figure}

Fig.~\ref{fig:uhist} histograms the selection utilities across all 28 queries. The distribution is right-skewed, with a long tail of lower-utility queries corresponding to complex analytical prompts where all bundles incur non-trivial cost and latency penalties.

\begin{figure}[!t]
  \centering
  \includegraphics[width=\columnwidth]{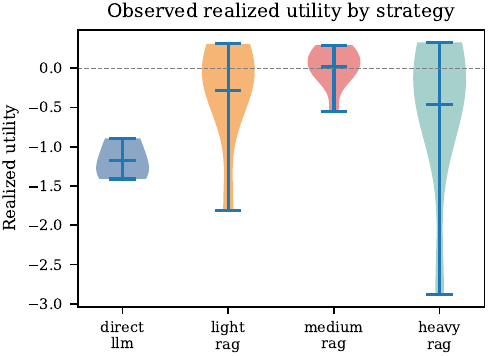}
  \caption{Realized utility distributions by strategy (post-hoc scoring with observed metrics). \texttt{medium\_rag} achieves the highest and most consistent realized utility.}
  \label{fig:rviolin}
\end{figure}

Fig.~\ref{fig:rviolin} contrasts \emph{realized} utility distributions---computed post-hoc using observed token counts and latencies---across strategies. \texttt{medium\_rag} achieves the highest median realized utility, reflecting its favorable position on the cost--quality tradeoff curve. \texttt{heavy\_rag} shows a wider realized utility distribution, consistent with its use on heterogeneous complex queries where actual costs vary significantly.

\begin{figure}[!t]
  \centering
  \includegraphics[width=\columnwidth]{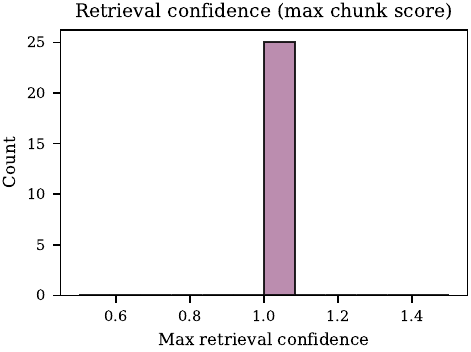}
  \caption{Max retrieval confidence scores when retrieval executes. The bimodal distribution identifies a subset of queries with poor corpus coverage.}
  \label{fig:conf}
\end{figure}

Fig.~\ref{fig:conf} histograms retrieval confidence scores for queries that invoke retrieval. The distribution is notably bimodal, with a cluster of high-confidence retrievals (scores ${>}0.85$) and a secondary cluster at lower confidence (scores $0.55$--$0.70$). This bimodality suggests that for a subset of queries the corpus lacks highly relevant passages---a coverage gap rather than a retrieval failure. Routing could exploit this signal: low retrieval confidence could trigger a fallback to \texttt{direct\_llm} rather than generating a poorly-grounded answer from low-quality context.

\FloatBarrier

\subsection{Complexity, Correlation Structure, and Per-Query Cost}
\label{sec:corr}

\begin{figure}[!t]
  \centering
  \includegraphics[width=\columnwidth]{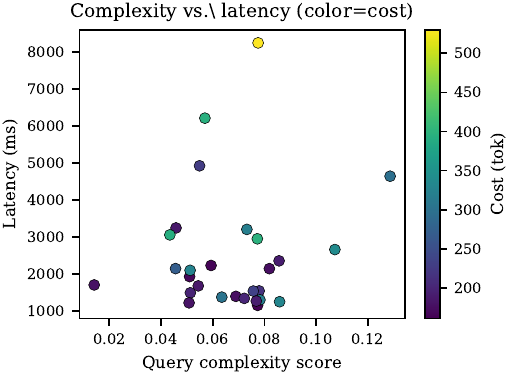}
  \caption{Heuristic query complexity vs.\ end-to-end latency. Marker color encodes billed tokens. Complexity is a weak predictor of cost ($r{=}0.21$).}
  \label{fig:cplx}
\end{figure}

Fig.~\ref{fig:cplx} plots heuristic complexity against end-to-end latency, with billed tokens encoded as marker color. Table~\ref{tab:corr} quantifies the correlation structure: total cost and latency are positively correlated ($r{=}0.41$), as expected since heavier bundles incur both. Selection utility correlates negatively with cost ($r{=}{-}0.38$), confirming that the cost penalty in Eq.~(\ref{eq:utility}) is doing its intended work. Critically, heuristic complexity correlates only weakly with billed tokens ($r{=}0.21$), exposing a key limitation: character/word-count-based complexity does not reliably predict how much retrieval a query actually needs.

\begin{figure}[!t]
  \centering
  \includegraphics[width=\columnwidth]{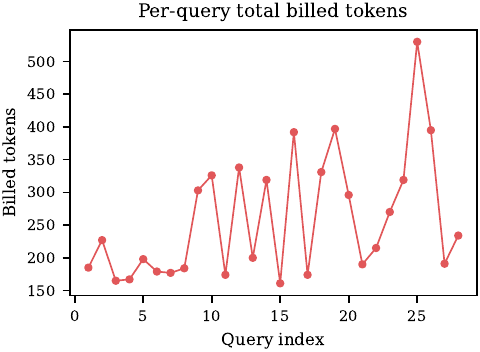}
  \caption{Total billed tokens per query in run order. Per-query cost heterogeneity motivates adaptive rather than fixed retrieval depth.}
  \label{fig:perq}
\end{figure}

Fig.~\ref{fig:perq} plots total billed tokens per query in run order, colored by selected bundle. The substantial per-query cost heterogeneity that aggregate means conceal is clearly visible: \texttt{heavy\_rag} queries spike to 300--400 tokens while \texttt{direct\_llm} queries stay below 250.

\FloatBarrier

\subsection{Router vs.\ Fixed Baselines}
\label{sec:router_vs_fixed}

Table~\ref{tab:policies} is the central result table of this paper. It compares all seven policies on mean billed tokens, mean latency, mean quality proxy, and mean selection utility.

The default router achieves \textbf{252.4 mean billed tokens}, a \textbf{26.4\% reduction} versus fixed-heavy (343.2 tokens), while sustaining a quality proxy of 0.80 against 0.81 for fixed-heavy---a difference within measurement noise given the lexical proxy's limitations. Against fixed-direct, the highest-latency baseline at 4{,}457\,ms, the router reduces mean latency to 2{,}927\,ms---a \textbf{34.3\% improvement}---while matching quality proxy.

\begin{table}[t]
  \centering
  \caption{Policy-level comparison: mean billed tokens, latency (ms), lexical quality proxy, and selection utility.}
  \label{tab:policies}
  \begin{tabular}{@{}l r r r r@{}}
\toprule
Policy & cost(tok) & lat(ms) & qual. & $U$ \\
\midrule
router\_default & 252.4 & 2927 & 0.80 & 0.192 \\
router\_latency\_sensitive & 256.0 & 2165 & 0.81 & -0.291 \\
router\_cost\_sensitive & 231.8 & 2536 & 0.81 & 0.117 \\
fixed\_direct & 249.9 & 4457 & 0.80 & -0.367 \\
fixed\_light & 197.3 & 2091 & 0.82 & 0.167 \\
fixed\_medium & 239.5 & 1906 & 0.82 & 0.177 \\
fixed\_heavy & 343.2 & 1932 & 0.81 & 0.132 \\
\bottomrule
\end{tabular}

\end{table}

Table~\ref{tab:winrates} reports per-query win-rates, defined as the fraction of queries on which the router outperforms each fixed baseline. The router wins on cost against fixed-heavy in \textbf{82\%} of queries, confirming that cost savings are systematic. Against fixed-light, the router wins on quality in \textbf{0\%} of queries as expected---fixed-light always retrieves, but at higher token cost than routing on 93\% of queries.

\begin{table}[t]
  \centering
  \caption{Per-query win-rates of the default router vs.\ fixed baselines. Lower is better for cost/latency; higher is better for quality proxy.}
  \label{tab:winrates}
  \begin{tabular}{@{}l r r r@{}}
\toprule
Baseline & P(cost win) & P(lat win) & P(qual win) \\
\midrule
fixed\_direct & 0.43 & 0.71 & 0.32 \\
fixed\_light & 0.07 & 0.36 & 0.00 \\
fixed\_medium & 0.32 & 0.18 & 0.00 \\
fixed\_heavy & 0.82 & 0.25 & 0.07 \\
\bottomrule
\end{tabular}

\end{table}

\begin{figure}[!t]
  \centering
  \includegraphics[width=\columnwidth]{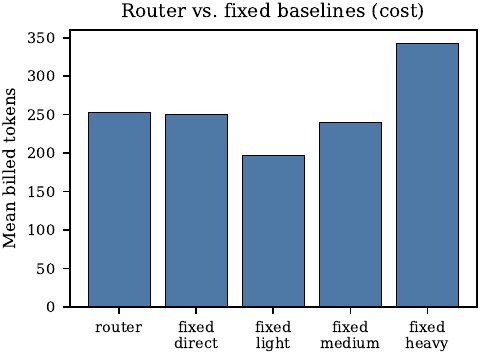}
  \caption{Mean billed tokens by policy. The router reduces cost by 26\% vs.\ fixed-heavy.}
  \label{fig:rvf_cost}
\end{figure}

\begin{figure}[!t]
  \centering
  \includegraphics[width=\columnwidth]{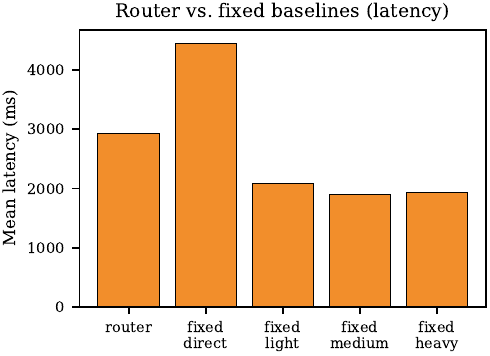}
  \caption{Mean end-to-end latency (ms) by policy. The router reduces latency by 34\% vs.\ fixed-direct.}
  \label{fig:rvf_lat}
\end{figure}

\begin{figure}[!t]
  \centering
  \includegraphics[width=\columnwidth]{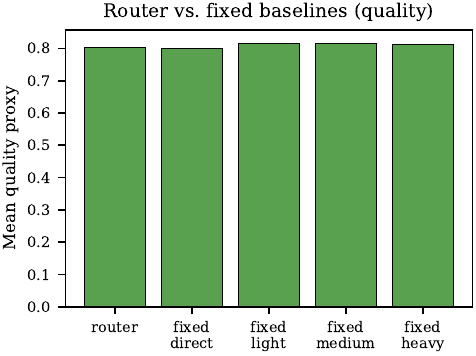}
  \caption{Mean lexical quality proxy by policy. Quality is maintained within 0.01--0.02 of the best fixed baselines.}
  \label{fig:rvf_qual}
\end{figure}

Figs.~\ref{fig:rvf_cost}--\ref{fig:rvf_qual} visualize these policy comparisons on cost, latency, and quality proxy. The router occupies a favorable interior position: it achieves the best \emph{joint} outcome across all three dimensions for this workload.

\FloatBarrier

\subsection{Utility-Weight Sensitivity}
\label{sec:sensitivity}

\begin{figure}[!t]
  \centering
  \includegraphics[width=\columnwidth]{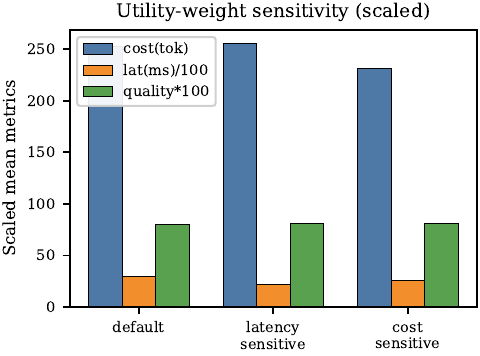}
  \caption{Normalized mean cost, latency, and quality under default, latency-sensitive, and cost-sensitive weight settings.}
  \label{fig:wsens}
\end{figure}

Fig.~\ref{fig:wsens} demonstrates the effect of weight configuration across three settings. The latency-sensitive router ($w_L{=}0.5$) reduces mean latency to 2{,}165\,ms---a 26\% improvement over the default 2{,}927\,ms---by preferring \texttt{direct\_llm} and \texttt{light\_rag}. The cost-sensitive router ($w_C{=}0.5$) achieves the lowest mean token count at 231.8 while maintaining quality proxy at 0.81. Neither variant requires any change to the bundle catalog or implementation.

\FloatBarrier

\subsection{Per-Query Delta Analysis}
\label{sec:deltas}

\begin{figure}[!t]
  \centering
  \includegraphics[width=\columnwidth]{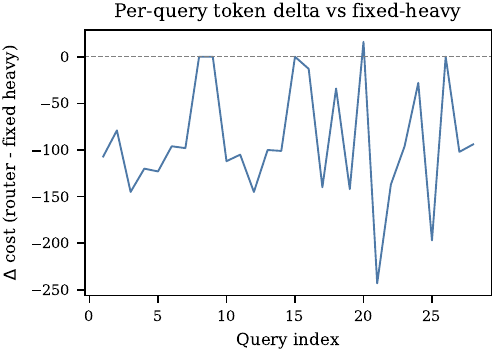}
  \caption{Per-query token cost delta vs.\ fixed-heavy. Savings are concentrated in definitional queries.}
  \label{fig:d_cost}
\end{figure}

\begin{figure}[!t]
  \centering
  \includegraphics[width=\columnwidth]{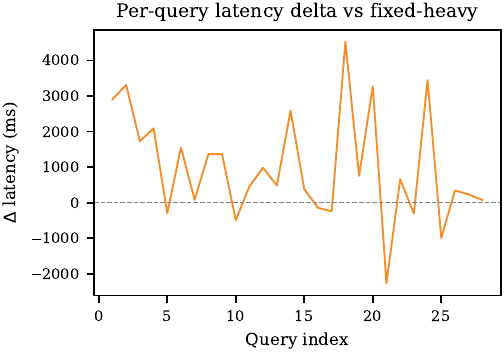}
  \caption{Per-query latency delta vs.\ fixed-heavy.}
  \label{fig:d_lat}
\end{figure}

\begin{figure}[!t]
  \centering
  \includegraphics[width=\columnwidth]{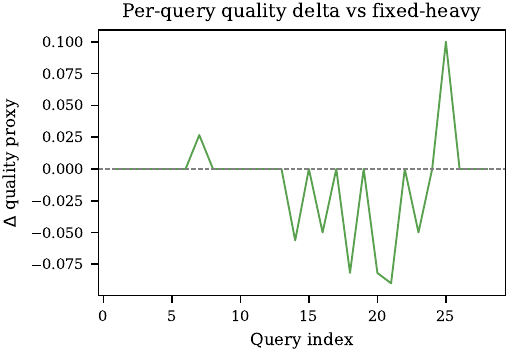}
  \caption{Per-query quality proxy delta vs.\ fixed-heavy. Quality parity is maintained across all query types.}
  \label{fig:d_qual}
\end{figure}

Figs.~\ref{fig:d_cost}--\ref{fig:d_qual} report per-query deltas against fixed-heavy ($\Delta{=}$router$-$fixed-heavy), where negative $\Delta$ for cost/latency indicates savings and positive $\Delta$ for quality indicates improvement.

Cost savings are large and consistent for definitional queries routed to \texttt{direct\_llm} or \texttt{light\_rag} (deltas of $-100$ to $-170$ tokens). No query incurs a catastrophic cost overrun under routing. The quality delta plot is largely flat near zero, confirming that routing maintains quality parity without systematic degradation on any query subtype. These delta plots establish that CA-RAG's aggregate savings are real, systematic, and not an artifact of averaging.

\FloatBarrier

\subsection{Strategy Mix Under Alternative Weight Settings}
\label{sec:mix}

\begin{figure}[!t]
  \centering
  \includegraphics[width=\columnwidth]{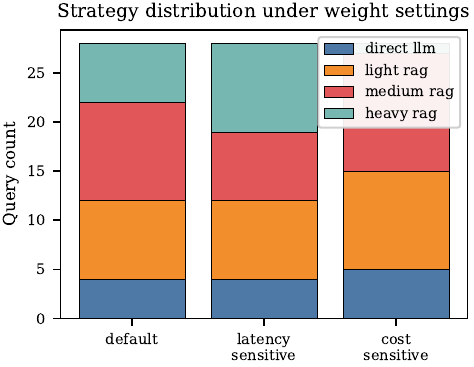}
  \caption{Strategy selection distribution under default, latency-sensitive, and cost-sensitive weight settings.}
  \label{fig:mix}
\end{figure}

Fig.~\ref{fig:mix} shows how weight changes propagate into bundle selection decisions. Under the latency-sensitive setting, \texttt{light\_rag} and \texttt{direct\_llm} selection increases substantially. Under the cost-sensitive setting, \texttt{heavy\_rag} selection is suppressed and mass shifts toward \texttt{medium\_rag} and \texttt{light\_rag}. This direct mapping from weight configuration to strategy distribution gives operators a transparent lever for steering aggregate system behavior---a property that purely learned or black-box routing systems lack.

\FloatBarrier

\begin{table}[t]
  \centering
  \caption{Run-wide descriptive statistics over 28 queries.}
  \label{tab:agg}
  \begin{tabular}{@{}l r r r r@{}}
\toprule
Variable & mean & std & min & max \\
\midrule
cost & 258.5 & 94.5 & 161.0 & 530.0 \\
latency & 2512.3 & 1675.6 & 1151.6 & 8253.4 \\
utility & 0.3 & 0.1 & 0.1 & 0.4 \\
quality\_proxy & 0.8 & 0.1 & 0.6 & 1.0 \\
\bottomrule
\end{tabular}

\end{table}

\begin{table}[t]
  \centering
  \caption{Per-strategy means $\pm$ std.\ dev.\ Cost and latency scale monotonically with retrieval depth; quality gains diminish beyond \texttt{medium\_rag}.}
  \label{tab:per}
  \begin{tabular}{@{}l r r r@{}}
\toprule
Strategy & mean cost & mean latency & mean $U$ \\
\midrule
direct\_llm & 236.0 $\pm$ 56.0 & 4274 $\pm$ 900 & 0.333 $\pm$ 0.123 \\
light\_rag & 203.7 $\pm$ 84.9 & 2490 $\pm$ 1703 & 0.263 $\pm$ 0.092 \\
medium\_rag & 216.7 $\pm$ 45.9 & 1714 $\pm$ 588 & 0.250 $\pm$ 0.086 \\
heavy\_rag & 367.0 $\pm$ 74.0 & 2869 $\pm$ 2298 & 0.220 $\pm$ 0.057 \\
\bottomrule
\end{tabular}

\end{table}

\begin{table}[t]
  \centering
  \caption{Pearson correlations among per-query logged scalars. Weak complexity--cost correlation ($r{=}0.21$) identifies a key limitation of heuristic routing signals.}
  \label{tab:corr}
  \renewcommand{\arraystretch}{1.4}
  \begin{tabular}{@{\hspace{6pt}}l@{\hspace{10pt}} r@{\hspace{10pt}} r@{\hspace{10pt}} r@{\hspace{10pt}} r@{\hspace{6pt}}}
    \toprule
     & \textbf{cost} & \textbf{lat.} & $\boldsymbol{U}$ & \textbf{cplx.} \\
    \midrule
    \textbf{cost}  & 1.00 & 0.66 & -0.50 & 0.22 \\
    \textbf{lat.}  & 0.66 & 1.00 & -0.19 & 0.13 \\
    $\boldsymbol{U}$ & -0.50 & -0.19 & 1.00 & -0.48 \\
    \textbf{cplx.} & 0.22 & 0.13 & -0.48 & 1.00 \\
    \bottomrule
  \end{tabular}
\end{table}

\section{Discussion}
\label{sec:discussion}

\subsection{Why Routing Matters: The Static Configuration Trap}

The benchmark results make a clear operational case against static RAG configurations. Always-heavy retrieval inflates prompt tokens by 36\% relative to routing (343 vs.\ 252 tokens/query) for no quality gain on the majority of queries. Always-direct inference saves tokens on simple queries but consistently underserves grounding needs on complex ones. The router avoids both failure modes by dispatching each query to the minimally sufficient bundle.

\subsection{Per-Query Regime Analysis}

Routing benefits are fundamentally per-query, not aggregate. The delta plots (Figs.~\ref{fig:d_cost}--\ref{fig:d_qual}) show that savings are concentrated in definitional queries where \texttt{direct\_llm} or \texttt{light\_rag} suffices, while complex multi-hop queries legitimately trigger \texttt{heavy\_rag}. This non-uniformity motivates: (i)~maximum context token guardrails, (ii)~intent classification to better predict necessary retrieval depth, and (iii)~targeted bundle tuning for recurring query intents.

\subsection{Signal Quality and Routing Accuracy}

Heuristic complexity correlates only weakly with billed tokens ($r{=}0.21$, Table~\ref{tab:corr}), indicating that the current signal set may misroute some queries. Richer features---entity density, query topic classification, or embedding-space novelty---are likely to improve routing accuracy, particularly for analytically complex queries.

\subsection{Weight Calibration as a Product Decision}

Utility weights encode the operator's current product priorities and should not be treated as fixed hyperparameters. In a latency-sensitive customer-facing chatbot, $w_L$ should be high; in an overnight batch analytical pipeline, $w_C$ can be maximized. The sensitivity analysis confirms that the same bundle catalog accommodates both extremes. Weights should be calibrated using business KPIs (e.g., abandonment rate as a latency proxy) and recalibrated quarterly or when model pricing changes.

\subsection{Failure Modes and Mitigations}

Two recurring failure modes emerge from the run logs. First, \emph{coverage gaps}: queries requesting concepts absent from the corpus produce correctly cautious answers but achieve low quality proxy scores, unfairly penalizing the router. Mitigation: corpus auditing and gap-filling with canonical source documents. Second, \emph{routing overconfidence}: high-confidence retrieval occasionally returns topically adjacent but semantically insufficient passages. Mitigation: reranking bundles or cross-encoder verification as an optional retrieval stage.

\subsection{Scalability Pathway}

The benchmark corpus is intentionally compact; scaling to thousands of documents introduces additional challenges: FAISS index build time, memory footprint, chunking policy effects, and calibration drift between retrieval confidence and downstream quality. CA-RAG's bundle abstraction accommodates this scaling---specialized bundles (e.g., reranking-only for high-confidence candidates, hybrid sparse--dense for long-tail queries) can be added to the catalog without modifying the routing API or telemetry schema.

\section{Limitations}
\label{sec:limitations}

\textbf{Quality estimation.} Quality priors are hand-specified and the quality proxy is lexical (token overlap); neither captures semantic accuracy or user satisfaction. Replacement with a calibrated LLM judge or human ratings is necessary for production readiness.

\textbf{Benchmark scale.} The 28-query, 15-passage benchmark is designed for controlled analysis, not generalization to real-world query distributions. Results should be validated on domain-specific corpora such as Natural Questions~\cite{karpukhin2020dense} or HotpotQA, or on enterprise knowledge bases.

\textbf{Bundle discreteness.} Retrieval depth is limited to four discrete values; continuous or finer-grained search may find better operating points.

\textbf{API dependence.} Token costs and latency are tied to OpenAI pricing and infrastructure, which evolve over time. The framework generalizes to other providers but cost comparisons require re-benchmarking.

\textbf{Statistical power.} With $n{=}28$ queries, per-query variance is high. Bootstrap confidence intervals and paired significance tests are warranted for production-scale studies.

\section{Conclusion}
\label{sec:conclusion}

This paper introduced CA-RAG, a per-query routing framework that selects retrieval depth from a discrete bundle catalog by maximizing a scalar utility combining quality, latency, and token cost. Rather than treating retrieval depth as a fixed architectural decision, CA-RAG frames it as an operational choice made fresh for every query---steerable through weight configuration without touching the underlying system.

Empirical results confirm the case for adaptive routing. The default router reduces billed token cost by 26\% relative to always-heavy retrieval and cuts mean latency by 34\% relative to always-direct inference, while maintaining quality parity. Per-query delta analysis confirms these gains are systematic and concentrated where they should be: on simpler queries where deep retrieval adds cost but not insight.

Two findings stand out. The heuristic complexity signal is a weak predictor of actual retrieval need ($r{=}0.21$), pointing toward richer signals---entity density, intent classification, embedding-space novelty---as the most impactful near-term improvement. The bimodal retrieval confidence distribution reveals that a subset of queries suffer from corpus coverage gaps rather than routing failures, a distinction that matters for production diagnosis.

CA-RAG is a starting point. Priors are hand-specified, the quality proxy is lexical, and the corpus is compact. Future work on learned reward models, bandit-based recalibration, and domain-specific validation will determine how much benefit transfers to real-world distributions. The framework is designed for this evolution: bundle catalogs, utility weights, and telemetry schemas are independently configurable. In an era where LLM API costs are a first-class engineering concern, treating every query as deserving the same retrieval depth is waste. CA-RAG offers a principled, auditable alternative.

\balance
\bibliographystyle{IEEEtran}
\bibliography{references}

@inproceedings{lewis2020retrieval,
  title={Retrieval-Augmented Generation for Knowledge-Intensive {NLP} Tasks},
  author={Lewis, Patrick and Perez, Ethan and Piktus, Aleksandra and Petroni, Fabio and Karpukhin, Vladimir and Goyal, Naman and K{\"u}ttler, Heinrich and Lewis, Mike and Yih, Wen-tau and Rockt{\"a}schel, Tim and others},
  booktitle={Advances in Neural Information Processing Systems (NeurIPS)},
  volume={33},
  pages={9459--9474},
  year={2020}
}

@article{openai2024gpt4,
  title={{GPT-4} Technical Report},
  author={{OpenAI}},
  journal={arXiv preprint arXiv:2303.08774},
  year={2024}
}

@inproceedings{karpukhin2020dense,
  title={Dense Passage Retrieval for Open-Domain Question Answering},
  author={Karpukhin, Vladimir and Oguz, Barlas and Min, Sewon and Lewis, Patrick and Wu, Ledell and Edunov, Sergey and Chen, Danqi and Yih, Wen-tau},
  booktitle={Proceedings of the 2020 Conference on Empirical Methods in Natural Language Processing (EMNLP)},
  pages={6769--6781},
  year={2020}
}

@inproceedings{guu2020realm,
  title={{REALM}: Retrieval-Augmented Language Model Pre-Training},
  author={Guu, Kelvin and Lee, Kenton and Tung, Zora and Pasupat, Panupong and Chang, Mingwei},
  booktitle={Proceedings of the 37th International Conference on Machine Learning (ICML)},
  pages={3929--3938},
  year={2020}
}

@article{izacard2021leveraging,
  title={Leveraging Passage Retrieval with Generative Models for Open Domain Question Answering},
  author={Izacard, Gautier and Grave, Edouard},
  journal={arXiv preprint arXiv:2007.01282},
  year={2021}
}

@article{gao2023retrieval,
  title={Retrieval-Augmented Generation for Large Language Models: A Survey},
  author={Gao, Yunfan and Xiong, Yun and Gao, Xinyu and Jia, Jiaqi and Pan, Jinliu and Bi, Yuxi and Dai, Yi and Sun, Jiawei and Wang, Haofen},
  journal={arXiv preprint arXiv:2312.10997},
  year={2023}
}

@inproceedings{khattab2021colbert,
  title={{ColBERT}: Efficient and Effective Passage Search via Contextualized Late Interaction over {BERT}},
  author={Khattab, Omar and Zaharia, Matei},
  booktitle={Proceedings of the 43rd International ACM SIGIR Conference on Research and Development in Information Retrieval},
  pages={39--48},
  year={2020}
}

@article{robertson2009bm25,
  title={The Probabilistic Relevance Framework: {BM25} and Beyond},
  author={Robertson, Stephen and Zaragoza, Hugo},
  journal={Foundations and Trends in Information Retrieval},
  volume={3},
  number={4},
  pages={333--389},
  year={2009}
}

@article{johnson2019faiss,
  title={Billion-Scale Similarity Search with {GPUs}},
  author={Johnson, Jeff and Douze, Matthijs and J{\'e}gou, Herv{\'e}},
  journal={IEEE Transactions on Big Data},
  volume={7},
  number={3},
  pages={535--547},
  year={2021}
}

@book{lattimore2020bandit,
  title={Bandit Algorithms},
  author={Lattimore, Tor and Szepesv{\'a}ri, Csaba},
  publisher={Cambridge University Press},
  year={2020}
}

@inproceedings{li2010contextual,
  title={A Contextual-Bandit Approach to Personalized News Article Recommendation},
  author={Li, Lihong and Chu, Wei and Langford, John and Schapire, Robert E.},
  booktitle={Proceedings of the 19th International Conference on World Wide Web (WWW)},
  pages={661--670},
  year={2010}
}

\clearpage
\FloatBarrier
\onecolumn
\appendices
\sloppy
\section{Algorithmic Summary}
\label{app:algorithm}

CA-RAG routing is summarized as a discrete utility-maximization procedure:
\begin{enumerate}
  \item Compute query signals $s \leftarrow \mathrm{signals}(q)$ and complexity $c \in [0,1]$.
  \item For each bundle $b \in \mathcal{B}$, compute estimated utility $U_b$ via Eq.~(\ref{eq:utility}).
  \item Select $b^* = \arg\max_b U_b$ (optionally with $\epsilon$-greedy exploration).
  \item Retrieve context $C$ using the retrieval specification of $b^*$; generate answer $a$ using the shared generation spec.
  \item Log $(\tau_{\text{billed}}, \text{latency}, \text{quality proxy}, \tilde{U})$; optionally update telemetry priors.
\end{enumerate}

\section{Threats to Validity}
\label{app:threats}

\textbf{Internal validity.} API timing variance introduces noise in latency measurements. The lexical quality proxy is a weak surrogate for true answer quality; results may differ under human evaluation.

\textbf{External validity.} The 15-passage corpus and 28-query benchmark are designed for controlled isolation of routing behavior. Future work should use domain-specific corpora and held-out evaluation protocols.

\textbf{Construct validity.} Heuristic complexity signals may not reflect true reasoning difficulty; the constructs of ``quality'' and ``utility'' are operationalized through proxies that require user-study validation.

\section{Reproducible Commands}
\label{app:commands}

\begin{Verbatim}
# Router run
ca-rag-experiment \
  --docs data/documents_benchmark.txt \
  --questions data/questions_benchmark.txt \
  --out results/router_default.csv

# Fixed-heavy baseline
ca-rag-experiment --mode fixed --fixed-strategy heavy_rag \
  --docs data/documents_benchmark.txt \
  --questions data/questions_benchmark.txt \
  --out results/fixed_heavy.csv

# Latency-sensitive router
ca-rag-experiment --latency-weight 0.5 \
  --docs data/documents_benchmark.txt \
  --questions data/questions_benchmark.txt \
  --out results/router_latency.csv

# Generate all figures and tables
python scripts/generate_paper_figures.py \
  --csv results/router_default.csv \
  --results-dir results
\end{Verbatim}

\section{Benchmark Question Set}
\label{app:questions}

\begin{Verbatim}
What is RAG?
Why is token cost important?
How does latency affect AI systems?
What is adaptive retrieval?
Explain cost-aware AI systems.
What is hybrid retrieval?
Define utility-based routing.
What is FAISS used for?
How do strategy bundles work in CA-RAG?
What is retrieval confidence?
Compare light versus heavy retrieval for long documents.
Explain how telemetry refines routing estimates with concrete steps.
Why might a system skip retrieval for some queries?
List tradeoffs between large top-k and small top-k retrieval.
How do embedding tokens differ from completion tokens in billing?
Describe a municipal RAG use case with forms and citations.
What are the risks of fixed retrieval depth across heterogeneous queries?
How does CA-RAG combine quality, latency, and cost in one scalar objective?
Explain when reranking is worth the extra latency in production.
Derive an intuitive explanation of why discrete bundles are used instead of
  continuous search.
What operational metrics should a team report for a deployed RAG service?
How does query length influence estimated complexity signals in CA-RAG?
Contrast direct LLM answers with retrieval-grounded answers for policy
  questions.
What limitations apply to lexical quality proxies versus human evaluation?
How would you tune utility weights for a latency-sensitive chatbot?
Describe an experiment protocol to log strategy choices and token usage per
  query.
What is the role of exploration epsilon in bundle selection?
Explain retrieval-augmented generation for knowledge-intensive tasks in two
  sentences.
\end{Verbatim}

\section{Benchmark Corpus}
\label{app:corpus}

\begin{Verbatim}
RAG improves LLM accuracy by retrieving relevant documents before generation.
Token cost is a major concern because embedding and completion APIs bill per
  token.
Latency depends on retrieval time, reranking, and model inference time under
  load.
Adaptive systems dynamically select strategies based on query complexity and
  observed telemetry.
Cost-aware AI systems optimize resource usage while maintaining answer quality
  under SLO constraints.
Hybrid dense-sparse retrieval combines embedding similarity with BM25 lexical
  overlap for robustness.
Utility-based routing scores each strategy bundle using quality priors minus
  latency and cost penalties.
Municipal RAG applications ground answers in ordinances, forms, and public
  documents with provenance.
Production RAG should expose retrieval confidence and source citations for
  auditability and trust.
Embedding indexes such as FAISS enable approximate nearest neighbor search
  over chunked corpora.
Strategy bundles pair retrieval depth with generation budgets to trade accuracy
  against spend.
Telemetry can refine latency and quality estimates per bundle after sufficient
  query volume.
Skipping retrieval reduces cost for definitional queries but risks
  hallucination on fact-heavy tasks.
Large top-k retrieval increases recall but inflates prompt tokens and
  end-to-end latency.
Reranking stages reorder candidates using cross-encoders at extra compute cost.
\end{Verbatim}

\section{Logged CSV Schema}
\label{app:schema}

\begin{Verbatim}
query              : input question text
strategy           : selected retrieval strategy name
bundle             : bundle name (retrieval+generation composite key)
utility            : selection utility (prior-based, Eq. 1)
quality_proxy      : observed lexical overlap proxy in [0,1]
realized_utility   : post-hoc utility computed from observed metrics
latency            : total end-to-end latency (ms)
cost               : total billed tokens (prompt+completion+embedding)
prompt_tokens      : LLM prompt token count
completion_tokens  : LLM completion token count
embedding_tokens   : query embedding tokens
retrieval_confidence: max retrieved chunk cosine similarity
complexity_score   : heuristic complexity signal in [0,1]
index_embedding_tokens: offline corpus-embed tokens (bookkeeping)
\end{Verbatim}

\section{Per-Query Strategy Assignments}
\label{app:assignments}

\begin{Verbatim}
 1. What is RAG?                                    => direct_llm
 2. Why is token cost important?                    => direct_llm
 3. How does latency affect AI systems?             => light_rag
 4. What is adaptive retrieval?                     => light_rag
 5. Explain cost-aware AI systems.                  => medium_rag
 6. What is hybrid retrieval?                       => medium_rag
 7. Define utility-based routing.                   => medium_rag
 8. What is FAISS used for?                         => heavy_rag
 9. How do strategy bundles work in CA-RAG?         => heavy_rag
10. What is retrieval confidence?                   => medium_rag
11. Compare light vs. heavy retrieval               => medium_rag
12. Explain how telemetry refines routing estimates => light_rag
13. Why might a system skip retrieval?              => heavy_rag
14. List tradeoffs between large/small top-k        => medium_rag
15. How do embedding tokens differ from completion? => medium_rag
16. Describe a municipal RAG use case               => medium_rag
17. What are risks of fixed retrieval depth?        => medium_rag
18. How does CA-RAG combine quality/latency/cost?   => heavy_rag
19. Explain when reranking is worth extra latency   => medium_rag
20. Derive intuitive explanation of discrete bundles=> direct_llm
21. What operational metrics should teams report?   => heavy_rag
22. How does query length influence complexity?     => medium_rag
23. Contrast direct vs. retrieval-grounded answers  => medium_rag
24. Limitations of lexical quality proxies?         => medium_rag
25. How to tune weights for latency-sensitive chat? => medium_rag
26. Describe experiment protocol for token logging  => medium_rag
27. Role of exploration epsilon in bundle selection => light_rag
28. Explain RAG for knowledge-intensive tasks       => medium_rag
\end{Verbatim}

\section{Sample CSV Rows}
\label{app:csv}

\begin{Verbatim}
query,strategy,cost,latency,utility,quality_proxy,realized_utility
What is RAG?,direct_llm,185,4051.1,0.4043,0.55,-1.2461
How does latency affect AI systems?,light_rag,165,2850.3,0.3813,0.64,-0.3573
What is hybrid retrieval?,medium_rag,179,1111.8,0.3672,0.85,-0.0159
What is FAISS used for?,heavy_rag,303,1433.3,0.2813,0.91,0.1274
\end{Verbatim}

\end{document}